\documentclass[twocolumn,prd,showpacs]{revtex4}
\usepackage{amssymb,amsmath}
\usepackage{amssymb}
\usepackage{amsmath}
\usepackage{graphicx,graphics}
\usepackage{color}
\usepackage{hyperref}
\begin{document}

\title{Localization of Matter Fields in the 6D Standing Wave Braneworld}

\author{Pavle Midodashvili${}^{1}$}
\email{pmidodashvili@yahoo.com}

\affiliation{${\ }^{(1)}
$Ilia State University, \\ 3/5 Kakutsa Cholokashvili Ave., Tbilisi 0162, Georgia\\}

\date{\today}

\begin{abstract}
We introduce new 6D standing wave braneworld model generated by gravity coupled to a phantom-like scalar field and investigate the problem of pure gravitational localization of matter fields. We show that in the case of increasing warp factor spin $0$, $1$, $1/2$ and $2$ fields are localized on the brane.
\end{abstract}


\maketitle


\section{Introduction}
The scenario where our world is associated with a brane embedded in a higher dimensional spacetime with non-factorizable geometry has attracted a lot of interest with the aim of solving several open questions in modern physics (see\cite{reviews-1,reviews-2,reviews-3,reviews-4} for reviews).

The braneworld models assume that all matter fields are localized on the brane, whereas gravity can propagate in the extra
dimensions. Recently we investigated the standing wave braneworld model in 5D spacetime  \cite{StandWave5D-1,StandWave5D-2,StandWave5D-3,StandWave5D-4,StandWave5D-5} and had shown that it provides universal gravitational trapping of zero modes of all kinds of matter fields in the case of rapid oscillations of standing waves in the bulk. The goal of this article is to generalize the model to 6D spacetime.

Although there exist a vast literature concerned to the localization of fields in 6D braneworld, there is not yet found
a universal trapping mechanism for all fields. In the existing 6D models with stationary exponentially warped spacetimes spin-$0$, spin-$1$ and spin-$2$ fields are localized on the brane with the decreasing warp factor, but spin-$1/2$ fields can be localized only with the increasing warp factor\cite{Od,Oda1}. There exist also 6D models with non-exponential warp factors providing gravitational localization of all kinds of bulk fields on the brane\cite{6D-0,6D-1,6D-2,6D-21,6D-3,6D-4,6D-5,6D-6}, however, these models require introduction of unnatural gravitational sources. There have been also considered models with time-dependent metrics and fields\cite{S-1,S-2,S-3,S-4}.

Here we introduce the non-stationary 6D braneworld model with gravity coupled to a phantom-like scalar field in the bulk, where generated bulk standing waves are bounded by the brane at the 2D extra space origin and the static part of the gravitational potential increasing at the extra space infinity. Then we explicitly show that this model provides universal gravitational trapping of zero modes of all kinds of matter fields in the case of rapid oscillations of bulk standing waves. It must be mentioned that analogous physical setup was considered in recent paper \cite{SSA}, which differs from our model in metric ansatz.

In the article in Section 2 we introduce the background solution of our model and some expressions for using in subsequent sections. Then, in Sections 3, 4, 5 and 6 we demonstrate existence of normalizable zero modes of spin-$0$, -$1$, -$1/2$ and -$2$ particles on the brane. Short final conclusions can be found in Section 7.

\section{Background Solution}

In 6D spacetime the Einstein equations with a bulk cosmological constant $\Lambda$ and stress-energy tensor $T_{AB}$ are
\begin{equation}\label{BS-EinsteinEquations}
{R_{AB}} - \frac{1}{2}{g_{AB}}R =    \Lambda{g_{AB}} + {k^2}{T_{AB}},
\end{equation}
where capital Latin indices refer to 6D spacetime  and 6D gravitational constant $k$ obeys the relation ${k^2} = 8\pi G = \frac{{8\pi }}{{{M^4}}}$  ($G$ and $M$ are the 6D Newton constant and the 6D Planck mass scale, respectively). For the two extra spatial dimensions we introduce polar coordinates ($r$,$\theta$), where $0\le r < \infty$ and $0 \le \theta < 2 \pi$.

In our physical setup there exists non-self-interacting phantom-like real scalar  field propagating in the bulk and having the following energy-momentum tensor
 \begin{equation}\label{BS-GhostEnergyMomentumTensor}
{T_{\rm{\text{Scal.field}}}}_{AB} =  - {\partial _A}\phi {\partial _B}\phi  + \frac{1}{2}{g_{AB}}{\partial ^C}\phi {\partial _C}\phi~.
\end{equation}
Using \eqref{BS-GhostEnergyMomentumTensor}, equations \eqref{BS-EinsteinEquations} can be rewritten in the form
\begin{equation}\label{BS-EinsteinEquations1}
R_{AB}=-\frac{1}{2}\Lambda g_{AB} - k^2 \partial_A \phi \partial_B \phi~.
\end{equation}
We look for the metric in the form
\begin{widetext}
\begin{equation}\label{BS-OutsideMetricAnsatz}
d{s^2} = {e^{2ar}}\left[ {d{t^2} - {e^{u\left( {t,r} \right)}}\left( {d{x^2} + d{y^2} + d{z^2}} \right)} \right] - d{r^2} - R_0^2{e^{2ar - 3u\left( {t,r} \right)}}d{\theta ^2},
\end{equation}
\end{widetext}
where curvature scale $a\neq0$ and $R_0>0$ are constants, and the function $u\left( {t,r} \right)$ depends only on time $t$ and on extra radial polar coordinate $r$. Then $u(t,r)=0$ this metric \textit{ansatz} coincides with the known solution \cite{Oda1}, that describes the model with a $3$-brane at the origin $r=0$ which is a 4D local string-like topological defect in the 6D spacetime. We also mention that the metric \textit{ansatz} \eqref{BS-OutsideMetricAnsatz} differs from that recently proposed in the article \cite{SSA}, our metric symmetrically considers $x$, $y$ and $z$ coordinates.

Taking into account symmetry properties of the metric \textit{ansatz} \eqref{BS-OutsideMetricAnsatz}, we assume  that phantom-like scalar field depends  only on time $t$ and extra coordinate $r$, i.e. $\phi=\phi\left(t,r\right)$. After a straightforward calculation, equations \eqref{BS-EinsteinEquations1} reduce to

\begin{eqnarray}\label{BS-EinstEqtnsSystem1}
3{\kern 1pt} {\left( {\frac{{\partial u}}{{\partial t}}} \right)^2} - 5{\kern 1pt} {a^2}{{\rm{e}}^{2{\kern 1pt} ar}} =   \frac{1}{2}\Lambda {e^{2ar}} +{k^2}{\left( {\frac{{\partial \phi }}{{\partial t}}} \right)^2}, \nonumber \\ \nonumber
3{\kern 1pt} \frac{{\partial u}}{{\partial t}}\frac{{\partial u}}{{\partial r}} = {k^2}\frac{{\partial \phi }}{{\partial t}}\frac{{\partial \phi }}{{\partial r}}, \\
\frac{{{\partial ^2}u}}{{\partial {t^2}}} - 10{\kern 1pt} {a^2}{{\rm{e}}^{2{\kern 1pt} ar}} - 5a{\kern 1pt} {{\rm{e}}^{2{\kern 1pt} ar}}\frac{{\partial u}}{{\partial r}} - {{\rm{e}}^{2{\kern 1pt} ar}}\frac{{{\partial ^2}u}}{{\partial {r^2}}} =   \Lambda {{\rm{e}}^{2{\kern 1pt} ar}}, \\ \nonumber
3{\mkern 1mu} {\left( {\frac{{\partial u}}{{\partial r}}} \right)^2} + 5{\mkern 1mu} {a^2} = -\frac{1}{2}\Lambda  + {k^2}{\left( {\frac{{\partial \phi }}{{\partial r}}} \right)^2}, \\ \nonumber
3{\mkern 1mu} \frac{{{\partial ^2}u}}{{\partial {t^2}}} + 10{\mkern 1mu} {a^2}{{\rm{e}}^{2{\kern 1pt} ar}} - 15a{\mkern 1mu} {{\rm{e}}^{2{\kern 1pt} ar}}\frac{{\partial u}}{{\partial r}} - 3{\mkern 1mu} {{\rm{e}}^{2{\kern 1pt} ar}}\frac{{{\partial ^2}u}}{{\partial {r^2}}} = \\ = -\Lambda {{\rm{e}}^{2{\kern 1pt} ar}}, \nonumber
\end{eqnarray}

from which we get
\begin{eqnarray}\label{BS-EinstEqtnsSystem2}
\Lambda = - 10a^2, \nonumber \\ \nonumber
3{\left( {\frac{{\partial u}}{{\partial t}}} \right)^2} = {k^2}{\left( {\frac{{\partial \phi }}{{\partial t}}} \right)^2}, \\
3\frac{{\partial u}}{{\partial t}}\frac{{\partial u}}{{\partial r}} = {k^2}\frac{{\partial \phi }}{{\partial t}}\frac{{\partial \phi }}{{\partial r}},\\ \nonumber
\frac{{{\partial ^2}u}}{{\partial {t^2}}} - 5a{\kern 1pt} {{\rm{e}}^{2{\kern 1pt} ar}}\frac{{\partial u}}{{\partial r}} - {{\rm{e}}^{2{\kern 1pt} ar}}\frac{{{\partial ^2}u}}{{\partial {r^2}}} = 0.
\end{eqnarray}

 First equation in system \eqref{BS-EinstEqtnsSystem2} fixes relation between bulk cosmological constant $\Lambda$ and the curvature scale $a$ in the exponential warp factor of the metric \eqref{BS-OutsideMetricAnsatz}. We see that the bulk cosmological constant $\Lambda$ must be negative.

 Taking into account the second and the third equations in \eqref{BS-EinstEqtnsSystem2}, we set the relation between metric function $u\left(t,r\right)$ and phantom-like scalar field \begin{math}\phi\left(t,r\right) \end{math} as

 \begin{equation}\label{BS-u-phi-relation}
\phi \left( {t,r} \right) = \sqrt {\frac{3}{{{k^2}}}} \,u\left( {t,r} \right).
 \end{equation}

Now we return to the last equation in \eqref{BS-EinstEqtnsSystem2} and, using separation of variables $u\left( {t,r} \right) = \tau \left( t \right)f\left( r \right)$, decouple it as follows
\begin{equation}\label{BS-Equations-tau-f}
\ddot \tau +\omega ^2\tau=0,~~~~f'' + 5af' + {\omega ^2}{e^{ - 2ar}}f = 0,
\end{equation}
where where $\omega>0$ is some real constant and overdots and primes denote derivatives with respect to $t$ and $r$, respectively.
The solution to the first equation in \eqref{BS-Equations-tau-f} is
\begin{equation}\label{BS-Solution-tau}
\tau \left( t \right) = {c_1}\sin \left( {\omega t} \right) + {c_2}\cos \left( {\omega t} \right),
\end{equation}
where $c_1$ and $c_2$ are some real constants.
To solve the second equation in  \eqref{BS-Equations-tau-f} we perform the following change of variables
\begin{equation}\label{BS-var-q}
z = \frac{\omega }{|a|}{e^{ - ar}},\,\,\,f\left( r \right) = {z^{\frac{5}{2}}}q\left( z \right),
\end{equation}
and it gets the form
\begin{equation}\label{BS-Equation-q}
\frac{{{d^2}q}}{{d{z^2}}} + \frac{1}{z}\frac{{dq}}{{dz}} + \left( {1 - \frac{{{{\left( {{5 \mathord{\left/
 {\vphantom {5 2}} \right.
 \kern-\nulldelimiterspace} 2}} \right)}^2}}}{{{z^2}}}} \right)q = 0~,
\end{equation}
with the solution
\begin{equation}\label{BS-Solution-q}
q\left( z \right) = {A_1}{J_{\frac{5}{2}}}\left( z \right) + {A_2}{Y_{\frac{5}{2}}}(z),
\end{equation}
where $A_{1}$ and $A_{2}$ are some real constants, and $J_{\frac{5}{2}}\left(z\right)$ and $Y_{\frac{5}{2}}\left(z\right)$ are $\frac{5}{2}$-order Bessel functions of first and second kind respectively. Taking into account change of variables \eqref{BS-var-q}, for $f(r)$ we get
\begin{eqnarray}\label{BS-Solution-f}
f\left( r \right) = {c_3}{e^{ - \frac{5}{2}ar}}{J_{\frac{5}{2}}}\left( {\frac{\omega }{|a|}{e^{ - ar}}} \right) + \\ \nonumber {c_4}{e^{ - \frac{5}{2}ar}}{Y_{\frac{5}{2}}}\left( {\frac{\omega }{|a|}{e^{ - ar}}} \right),
\end{eqnarray}
where $c_3$ and $c_4$ are some real constants.

Imposing on $u(t,r)$ the following boundary condition at the extra space infinity
\begin{equation}\label{BS-BoundaryConditions}
u\left( {t, + \infty } \right) = 0,
\end{equation}
and taking the case of increasing warp factor, i.e. $a>0$, from \eqref{BS-Solution-tau} and \eqref{BS-Solution-f} we get
\begin{equation}\label{BS-Function-u}
u(t,r) = \sin \left( {\omega t} \right)B(r)~,
\end{equation}
with
\begin{equation}\label{BS-Function-B}
B(r)=B_0{e^{ - \frac{5}{2}ar}}{J_{\frac{5}{2}}}\left( {\frac{\omega }{{a}}{e^{ - ar}}} \right)~,
\end{equation}
where $B_0$ is some real constant.

 The ghost-like field $\phi(t,r)$ and the metric oscillations via the metric oscillatory function $u(t,r)$ must be unobservable on the brane located at $r=0$ in the extra space. According to \eqref{BS-u-phi-relation}, \eqref{BS-Function-u} and \eqref{BS-Function-B} we can fulfil the requirement by imposing the boundary condition on the Bessel function
\begin{equation}\label{BS-BoundaryCondition-Y}
\left. {{J_{\frac{5}{2}}}\left( {\frac{\omega }{{a}}{e^{ - ar}}} \right)} \right|_{r = 0 }={{J_{\frac{5}{2}}}\left( {\frac{\omega }{{a}}} \right)} = 0~.
\end{equation}
Taking into account the oscillatory character of the Bessel function $J_{\frac{5}{2}}$ and denoting by $Z_n$ its $n$-th zero, this condition can be written in the form
\begin{equation}\label{BS-Quantization-omega-a-epsilon}
\frac{\omega }{{a}} = {Z_n}~,
\end{equation}
 which quantizes the standing wave oscillation frequency $\omega$ in terms or curvature scale $a$. Choosing some value of $n$ in \eqref{BS-Quantization-omega-a-epsilon}, the function $B(r)$ gets the form
 \begin{equation}\label{BS-Function-B(r)-withZn}
 B(r) = {B_0}{e^{ - \frac{5}{2}ar}}{J_{\frac{5}{2}}}\left( {{Z_n}\,{e^{ - ar}}} \right),
 \end{equation}
and it's easy to see that in the 2D extra space we will have $(n-1)$ concentric circles (all of them sharing the same center coinciding with the 2D extra space origin) where the oscillatory function ${J_{\frac{5}{2}}}\left( {{Z_n}\,{e^{ - ar}}} \right)$ vanishes. The radiuses $r_i$ of these circles satisfy the relation
\begin{equation}\label{BS-ZeroCircle-Radiuses}
Z_n{e^{ - ar_i }} = {Z_i}~,
\end{equation}
where $i =0, 1, ..., n-1$. These circles are the nodes of the circular standing wave in the 2D extra space and can be considered as the circular islands where matter particles can be bound. It's obvious that the metric function $u(t,r)$ \eqref{BS-Function-B} and scalar field $\phi(t,r)$ \eqref{BS-u-phi-relation} also vanish at the circles. In what follows in \eqref{BS-Quantization-omega-a-epsilon} we assume n=1, i.e. we choose the first zero of the function $J_{\frac{5}{2}}$
\begin{equation}\label{BS-BesselJZeros(1)}
Z_1 \approx 5.763.
\end{equation}
In this case circular standing wave has only two nodes in the extra space: at the origin $r=0$ and at the infinity $r=+\infty$.

In the subsequent sections we investigate various matter field equations. The metric oscillatory function \eqref{BS-Function-u}  enters the equations via exponents $e^{bu(t,r)}$, with $b$ denoting some real constant. To solve the localization problem we assume that the standing wave frequency $\omega$ is much larger than the frequencies corresponding to the energies of the particles localized on the string-like brane, and in the matter field equations we perform time averaging of the oscillating exponents. Denoting the time average $\left \langle e^{bu(t,r)}\right \rangle$ by $K_{b}(r)$  and using the results of our previous papers \cite{StandWave5D-2,StandWave5D-3,StandWave5D-5}, we can write the following useful expressions
\begin{eqnarray}\label{BS-TimeAverages}
{K_{b}}(r) = \left\langle {{e^{bu(t,r)}}} \right\rangle  = \left\langle {{e^{ - bu(t,r)}}} \right\rangle = \nonumber \\  ={K_{-b}}(r)= {I_0}(bB(r))~,\\ \nonumber
\left\langle u \right\rangle  = \left\langle {\frac{{\partial u}}{{\partial r}}} \right\rangle  = \left\langle {\frac{{\partial u}}{{\partial t}}} \right\rangle  = \left\langle {\frac{{\partial \left( {{e^{bu}}} \right)}}{{\partial t}}} \right\rangle  = 0~,
\end{eqnarray}
where $I_0$ is the modified Bessel function of order zero and the function $B(r)$ is defined by \eqref{BS-Function-B(r)-withZn} with $n=1$. In what follows we also use the asymptotic expansions of the function ${K_{b}}(r)$ at the origin and infinity in the extra space
\begin{widetext}
\begin{eqnarray}\label{BS-K_b(r)-Asymptotics}
{\left. {{K_b}\left( r \right)} \right|_{r \to 0}} &=& 1 + \frac{{{b^2}{B_0}^2{{\left[ {\sin \left( {{Z_1}} \right) - {Z_1}\cos \left( {{Z_1}} \right)} \right]}^2}}}{{2\pi {Z_1}}}\,\,{a^2}{r^2} + O\left( {{a^3}{r^3}} \right),\\
{\left. {{K_b}\left( r \right)} \right|_{r \to  + \infty }} &=& 1 + \frac{{{b^2}{B_0}^2{Z_1}^5}}{{450\pi }}\,\,{e^{ - 10ar}} + O\left( {{e^{ - 12ar}}} \right) ,\nonumber
\end{eqnarray}
\end{widetext}
with $Z_1$ defined by \eqref{BS-BesselJZeros(1)}.

\section{Localization of Scalar Fields}

We start with the problem of localization of massless scalar fields defined by the 6D action
\begin{equation} \label{SF-action}
{S_0} =  - \frac{1}{2}\int {{d^6}x\sqrt { - g}~ {g^{MN}}{\partial _M}\Phi {\partial _N}\Phi } ~,
\end{equation}
where the determinant for our \textit{ansatz} \eqref{BS-OutsideMetricAnsatz} has the following form
\begin{equation} \label{SF-determinant}
 \sqrt {- g} =R_0 e^{5ar}~.
\end{equation}
Performing time averaging of the oscillating functions in the corresponding Klein-Gordon equation,
\begin{equation}\label{SF-KGEquation1}
\frac{1}{{\sqrt {- g} }}{\partial _M}\left( {\sqrt {- g} ~{g^{MN}}{\partial _N}\Phi } \right) = 0~,
\end{equation}
we get
\begin{eqnarray}\label{SF-KGEquation2}
\left[ {\partial _t^2 -  K_1\left( {\partial _x^2 + \partial _y^2 + \partial _z^2} \right) - \frac{1}{{R_0^2}}K_3\partial _\theta ^2} \right]\Phi  = \\ \nonumber = {e^{2ar}}\left[ {\partial _r^2 + 5a{\partial _r}} \right]\Phi ~,
\end{eqnarray}
where functions $K_1(r)$ and $K_3(r)$ are time averages defined by \eqref{BS-TimeAverages}.
We look for the solution to this equation in the form
\begin{equation}\label{SF-Solution}
\Phi \left( {t,x,y,z,r,\theta } \right) = {e^{i\left(Et - {p_n}x^n \right)}}\sum\limits_{l,m} {\phi_{m} (r){e^{il\theta }}}~,
\end{equation}
with ${p_n}x^n={p_x}x + {p_y}y + {p_z}z$ and $\phi_m(r)$ obeying the following equation
\begin{widetext}
\begin{equation}\label{SF-Equation-phi-1}
\left[ {\frac{{{d^2}}}{{d{r^2}}} + 5a\frac{d}{{dr}} + {e^{ - 2ar}}\left( {{E^2} - {p^2}{K_1} - \frac{{{l^2}}}{{R_0^2}}{K_3}} \right)} \right]{\phi _m}(r) = 0~,
\end{equation}
\end{widetext}
where $p^2 = p_x^2 + p_y^2 + p_z^2$. On the brane, where $u\approx 0$, the parameters $E$, $p_x$, $p_y$ and $p_z$ can be considered as energy and momentum components along 4D.

Taking into account that the energy and momentum of 4D massless scalar field obey the dispersion relation
\begin{equation}
E^2 = p_x^2 + p_y^2 + p_z^2~,
\end{equation}
for $\phi_0(r)$ of the s-wave ($l=0$) zero mode wave function  we get the equation
\begin{equation}\label{SF-Equation-rho}
\left[ {\frac{{{d^2}}}{{d{r^2}}} + 5a\frac{d}{{dr}} - {p^2}{e^{ - 2ar}}\left( {K_1 - 1} \right)} \right]{\phi _0}(r) = 0~.
\end{equation}
We put \eqref{SF-Equation-rho} into the form of an analogue non-relativistic quantum mechanical problem by making the change
\begin{equation} \label{SF-varsigma}
\phi_0 (r) = {e^{ - \frac{5}{2}ar}}\sigma_0 (r)~,
\end{equation}
and for $\sigma_0 (r)$ we find
\begin{equation}\label{SF-Equation-varsigma}
\left[ {\frac{{{d^2}}}{{d{r^2}}} - U_0(r)} \right]{\sigma _0} = 0~,
\end{equation}
where the function
\begin{equation}\label{SF-U(r)}
U_0(r) = \frac{{25}}{4}{a^2} + {p^2}{e^{ - 2ar}}\left( {K_1 - 1} \right)
\end{equation}
is the analog of non-relativistic potential. Figure 1 shows behaviour of $U_0(r)$.


\begin{figure}[h]
\begin{center}
\includegraphics[width=0.3\textwidth]{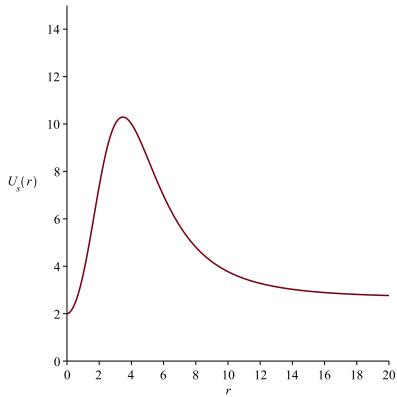}
\caption{The effective bulk potential \eqref{SF-U(r)}.}
\end{center}
\end{figure}


We explore (\ref{SF-Equation-varsigma}) in two limiting regions, far from and close to the brane.
Using \eqref{BS-K_b(r)-Asymptotics}, far from the brane, i.e. $r \to +\infty$, the equation (\ref{SF-Equation-varsigma}) obtains the form
\begin{equation}\label{SF-Equation-varsigma-infinity}
\left[ {\frac{{{d^2}}}{{d{r^2}}} - \frac{{25}}{4}{a^2}} \right]{\sigma _0} = 0~,
\end{equation}
with the general solution
\begin{equation}\label{SF-Solution-varsigma-infinity}
\sigma_0 \left( r \right) = {C_1}{e^{{{-5ar} \mathord{\left/
 {\vphantom {{5ar} 2}} \right.
 \kern-\nulldelimiterspace} 2}}} + {C_2}{e^{{{ 5ar} \mathord{\left/
 {\vphantom {{ - 5ar} 2}} \right.
 \kern-\nulldelimiterspace} 2}}},
\end{equation}
where $C_1$ and $C_2$ are some constants.
Taking into account \eqref{SF-varsigma}, the extra dimension factor $\phi_0 (r)$ of the scalar field gets the following form
\begin{equation}\label{SF-rho-infinity}
\left.\phi_0 \left( r \right)\right|_{r \to + \infty } =  {C_1}{e^{ - 5ar}}+{C_2}.
\end{equation}
To have normalizable zero mode, at the infinity we impose the boundary condition
\begin{equation}\label{SF-rho-BoundaryCondition-infinity}
\left.\phi_0 \left( r \right)\right|_{r \to + \infty } =0.
\end{equation}
So we set $C_2=0$ in \eqref{SF-rho-infinity} and get the following result

\begin{equation}\label{SF-rho-infinity1}
\left.\phi_0 \left( r \right)\right|_{r \to + \infty } = {C_1}{e^{ - 5ar}}~.
\end{equation}

Taking into account \eqref{BS-K_b(r)-Asymptotics}, close to the brane, $r \to 0$, the equation \eqref{SF-Equation-varsigma} again reduces to \eqref{SF-Equation-varsigma-infinity}. Resemblance of the equations far from and close to the brane is not surprising since according to \eqref{BS-K_b(r)-Asymptotics} the second term of $U(r)$ in \eqref{SF-U(r)} vanishes in both regions (these two regions correspond to the nodes of the standing waves). Therefore for the extra dimension factor $\phi_0 (r)$ of the scalar field zero mode we again get

\begin{equation}\label{SF-rho-origin}
\left.\phi_0 \left( r \right)\right|_{r \to 0 } = {C_3}{e^{ - 5ar}}+ {C_4}~,
\end{equation}
where $C_3$ and $C_4$ are some constants. At the origin $r=0$ we impose boundary condition

\begin{equation}\label{SF-rho-BoundaryCondition-origin}
\left. {\frac{{d{\phi _0}\left( r \right)}}{{dr}}} \right|_{r = 0} = 0~,
\end{equation}
and get the following result
\begin{equation}\label{SF-rho-origin1}
\left.\phi_0 \left( r \right)\right|_{r \to 0 } = {C_3}\left ( 5a + {e^{ - 5ar}}\right )~.
\end{equation}

So $\phi_0 (r)$ has a maximum on the brane and falls off at the infinity as $e^{-5ar}$. In the scalar field action \eqref{SF-action} the determinant \eqref{SF-determinant} and the metric tensor with upper indices give the total exponential factor $e^{3ar}$, which obviously increases for $a > 0$. But the extra part of wave function \eqref{SF-rho-infinity1} contains the exponentially decreasing factor $e^{-5ar}$. So, for such extra dimension factor the integral over $r$ in the action \eqref{SF-action} is convergent, i.e. 4D scalar fields are localized on the brane.


\section{Localization of Vector Fields}

For simplicity we consider only the $U(1)$ vector field, the generalization to the case of non-Abelian gauge fields is straightforward. The action of vector field is:
\begin{equation}\label{VectorAction}
S_1 = - \frac{1}{4}\int d^6x\sqrt {-g}~ g^{MN}g^{PR}F_{MP}F_{NR}~,
\end{equation}
where
\begin{equation} \label{F}
F_{MP} = \partial _M A_P - \partial _P A_M
\end{equation}
is the 6D vector field tensor, and the determinant $g$ of the metric is defined by \eqref{SF-determinant}.

The action (\ref{VectorAction}) gives the system of six equations
\begin{equation}\label{VectorFieldEquations}
\frac{1}{\sqrt{- g} }\partial _M\left( \sqrt{- g} ~g^{MN}g^{PR}F_{NR} \right) = 0 ~,
\end{equation}
which for our metric \eqref{BS-OutsideMetricAnsatz} have the following explicit form
\begin{widetext}
\begin{eqnarray}\label{VectorFieldEquations1}
{e^{ar - u}}\left( {{\partial _x}{F_{xt}} + {\partial _y}{F_{yt}} + {\partial _z}{F_{zt}}} \right) + {\partial _r}\left( {{e^{3ar}}{F_{rt}}} \right) + {{R_0^{-2}}}{e^{ar + 3u}}{\partial _\theta }{F_{\theta t}} = 0, \nonumber \\
{\partial _t}\left( {{e^{ar - u}}{F_{tx}}} \right) - {\partial _y}\left( {{e^{ar - 2u}}{F_{yx}}} \right) - {\partial _z}\left( {{e^{ar - 2u}}{F_{zx}}} \right) - {\partial _r}\left( {{e^{3ar - u}}{F_{rx}}} \right) - {{R_0^{-2}}}{e^{ar + 2u}}{\partial _\theta }{F_{\theta x}} = 0, \nonumber \\
{\partial _t}\left( {{e^{ar - u}}{F_{ty}}} \right) - {\partial _x}\left( {{e^{ar - 2u}}{F_{xy}}} \right) - {\partial _z}\left( {{e^{ar - 2u}}{F_{zy}}} \right) - {\partial _r}\left( {{e^{3ar - u}}{F_{ry}}} \right) - {{R_0^{-2}}}{e^{ar + 2u}}{\partial _\theta }{F_{\theta y}} = 0,\\
{\partial _t}\left( {{e^{ar - u}}{F_{tz}}} \right) - {\partial _x}\left( {{e^{ar - 2u}}{F_{xz}}} \right) - {\partial _y}\left( {{e^{ar - 2u}}{F_{yz}}} \right) - {\partial _r}\left( {{e^{3ar - u}}{F_{rz}}} \right) - {{R_0^{-2}}}{e^{ar + 2u}}{\partial _\theta }{F_{\theta z}} = 0, \nonumber \\
{\partial _t}\left( {{e^{3ar}}{F_{tr}}} \right) - {\partial _x}\left( {{e^{3ar - u}}{F_{xr}}} \right) - {\partial _y}\left( {{e^{3ar - u}}{F_{yr}}} \right) - {\partial _z}\left( {{e^{3ar - u}}{F_{zr}}} \right) - {{R_0^{-2}}}{e^{3ar + 3u}}{\partial _\theta }{F_{\theta r}} = 0, \nonumber \\
{\partial _t}\left( {{e^{ar + 3u}}{F_{t\theta }}} \right) - {\partial _x}\left( {{e^{ar + 2u}}{F_{x\theta }}} \right) - {\partial _y}\left( {{e^{ar + 2u}}{F_{y\theta }}} \right) - {\partial _z}\left( {{e^{ar + 2u}}{F_{z\theta }}} \right) - {\partial _r}\left( {{e^{3ar + 3u}}{F_{r\theta }}} \right) = 0. \nonumber
\end{eqnarray}
\end{widetext}
We look for the solution to the system (\ref{VectorFieldEquations1}) in the form:
\begin{eqnarray}\label{VectorFieldDecomposition}
{A_t}({x^C}) = {a_t}({x^\nu })\sum\limits_{l,m} {{\rho _m}(r){e^{il\theta }}} \nonumber ,\\ {A_k}({x^C}) = {e^{u(t,r)}}{a_k}({x^\nu })\sum\limits_{l,m} {{\rho _m}(r){e^{il\theta }}} ,\\
{A_r}({x^C}){\rm{ }} = {a_r}({x^\nu })\sum\limits_{l,m} {{\xi _m}(r){e^{il\theta }}} , \nonumber \\ {A_\theta }({x^C}){\rm{ }} = 0~, \nonumber
\end{eqnarray}
where index $k$ runs $x$, $y$ and $z$, $u(t, r)$ is the oscillatory metric function \eqref{BS-Function-u}, and $a_{\mu}(x^\nu)$ denote the components of 4D vector potential (Greek letters are used for 4D indices) that obey 4D Lorenz-like gauge condition
\begin{equation}\label{4D-LorenzCondition}
{\eta ^{\alpha \beta }}{\partial _\alpha }{a_\beta }({x^\nu }) = 0
\end{equation}
($\eta _{\alpha \beta }$ denote the metric of 4D Minkowski spacetime). In fact, equation \eqref{4D-LorenzCondition} together with the last expression in \eqref{VectorFieldDecomposition} can be considered as the full set of gauge conditions imposed on the vector field $A_M(x^C)$.

In the case when the frequencies of bulk standing waves $\omega$ is much larger than frequencies associated with the energies of the particles on the brane,
\begin{equation}
\omega \gg E~,
\end{equation}
we assume the existence of localized flat 4D vector waves,
\begin{equation} \label{Factors}
a_\mu\left(x^\nu\right) \sim \varepsilon_\mu e^{i(Et - p_x x - p_y y - p_z z)} ~,
\end{equation}
where $E$, $p_x$, $p_y$, $p_z$ are components of energy-momentum along the brane. So putting \eqref{VectorFieldDecomposition} into \eqref{VectorFieldEquations1} and performing time averaging we get the following system of equations
\begin{widetext}
\begin{eqnarray}
\label{VectorFieldEquation-(tt)}
\left( {{\partial _r}\left( {{e^{3ar}}{\partial _r}} \right) - {e^{ar}}\left( {{K_1}{\partial ^\mu }{\partial _\mu } - \left( {{K_1} - 1} \right)\partial _t^2 + \frac{{{l^2}}}{{R_0^2}}{K_3}} \right)} \right){a_t}{\rho _m} = {\partial _r}\left( {{e^{3ar}}{\xi _m}} \right){\partial _t}{a_r}~,\\
\label{VectorFieldEquation-(xx-yy-zz)}
\left( {{\partial _r}\left( {{e^{3ar}}{\partial _r}} \right) - {e^{ar}}\left( {{K_1}{\partial ^\mu }{\partial _\mu } - \left( {{K_1} - 1} \right)\partial _t^2 + \frac{{{l^2}}}{{R_0^2}}{K_3}} \right)} \right){a_i}{\rho _m} = {K_1}{\partial _r}\left( {{e^{3ar}}{\xi _m}} \right){\partial _i}{a_r}~,\\
\label{VectorFieldEquation-(rr)}
\left( {{\partial ^\mu }{\partial _\mu } + \left( {{K_1} - 1} \right)\partial _t^2 + \frac{{{l^2}}}{{R_0^2}}{K_4}} \right){a_r}{\xi _m} = 0~,\\
\label{VectorFieldEquation-(ThetaTheta)}
{\partial _r}\left( {{e^{3ar}}{\xi _m}} \right) = 0~,
\end{eqnarray}
\end{widetext}
where in \eqref{VectorFieldEquation-(xx-yy-zz)} index $i$ runs values $x$, $y$ and $z$, and the functions $K_1(r)$, $K_3(r)$ and $K_4(r)$ are time averages defined by \eqref{BS-TimeAverages}. For the $s$-wave ($l=0$) zero mode wave function this system reduces to
\begin{widetext}
\begin{eqnarray}
\label{VectorFieldEquation-(tt)zero}
\left( {{\partial _r}\left( {{e^{3ar}}{\partial _r}} \right) - {e^{ar}}{p^2}\left( {{K_1} - 1} \right)} \right){a_t}{\rho _0} = {\partial _r}\left( {{e^{3ar}}{\xi _0}} \right){\partial _t}a_r~,\\
\label{VectorFieldEquation-(xx-yy-zz)zero}
\left( {{\partial _r}\left( {{e^{3ar}}{\partial _r}} \right) - {e^{ar}}{p^2}\left( {{K_1} - 1} \right)} \right){a_i}{\rho _0} = {K_1}{\partial _r}\left( {{e^{3ar}}{\xi _0}} \right){\partial _i}{a_r}~,\\
\label{VectorFieldEquation-(rr)zero}
{p^2}\left( {{K_1} - 1} \right){a_r}{\xi _0} = 0~,\\
\label{VectorFieldEquation-(ThetaTheta)zero}
{\partial _r}\left( {{e^{3ar}}{\xi _0}} \right) = 0~,
\end{eqnarray}
\end{widetext}
with $p^2=p_x^2+p_y^2+p_z^2$.
According to \eqref{VectorFieldEquation-(rr)zero} and \eqref{VectorFieldEquation-(ThetaTheta)zero} we set  $\xi _0(r)=C_1e^{-3ar}$ and $a_{r} (x^{\nu})=0$, then equations \eqref{VectorFieldEquation-(tt)zero} and \eqref{VectorFieldEquation-(xx-yy-zz)zero} reduce to the following single equation
\begin{equation}\label{VF-Equation-rho}
\left( {\frac{d}{{dr}}\left( {{e^{3ar}}\frac{d}{{dr}}} \right) - {e^{ar}}{p^2}\left( {{K_1} - 1} \right)} \right){\rho _0}(r) = 0~.
\end{equation}
So the zero mode solution has the following 6D form
\begin{eqnarray}\label{VF-ZeroModeSolution-6D}
{A_t}({x^C}){\rm{ }} &=& {a_t}({x^\nu }){\rho _0}(r), \nonumber  \\ {A_i}({x^C}){\rm{ }} &=& {e^{u(t,r)}}{a_i}({x^\nu }){\rho _0}(r), \\ \nonumber {A_r}({x^C}){\rm{ }} &=& 0, \\ {A_\theta }({x^C}){\rm{ }} &=& 0, \nonumber
\end{eqnarray}
where 4D factors ${a_\mu}({x^\nu })$ have form defined by \eqref{Factors}, and the function $\rho_0(r)$ obeys the equation \eqref{VF-Equation-rho}.

By making the change
\begin{equation}\label{VF-varupsilon}
{\rho _0}(r) = e^{ - 3ar/2}\upsilon _0 (r),
\end{equation}
we rewrite \eqref{VF-Equation-rho} into the form of non-relativistic quantum mechanical problem
\begin{equation}\label{VF-Equation-varupsilon}
\left[ {\frac{{{d^2}}}{{d{r^2}}} - U_1(r)} \right]{\upsilon _0} = 0~,
\end{equation}
where the potential
\begin{equation}\label{VF-U(r)}
U_1(r) = \frac{{9}}{4}{a^2} + {p^2}{e^{ - 2ar}}\left( {K_1 - 1} \right)
\end{equation}
differs from the analogous potential for scalar fields \eqref{SF-U(r)} (see Figure 1) only in constant factor in the first term. So, using the same arguments as in the previous section, the solutions to \eqref{VF-Equation-varupsilon} in the two limiting regions - far from and close to the brane - are
\begin{eqnarray}\label{VS-Solution-varupsilon-infinity}
\left.\upsilon _0\left( r \right)\right|_{r \to + \infty } &=& {C_1}{e^{-3ar/2}} + {C_2}{e^{3ar/2}},\\ \nonumber
\left.\upsilon _0\left( r \right)\right|_{r \to 0} &=& {C_3}{e^{-3ar/2}} + {C_4}{e^{3ar/2}},
\end{eqnarray}
where $C_1$, $C_2$, $C_3$ and $C_4$ are again some constants. Taking into account \eqref{VF-varupsilon} and imposing boundary conditions analogues to \eqref{SF-rho-BoundaryCondition-infinity} and \eqref{SF-rho-BoundaryCondition-origin}, the extra dimension factor $\rho_0(r)$ of the vector field zero mode will have the following asymptotic forms
\begin{eqnarray}\label{VF-Solution-rho-asymptotics}
\left.\rho _0\left( r \right)\right|_{r \to + \infty } &=& {C_1}{e^{ - 3ar}}, \\ \nonumber
\left.\rho _0\left( r \right)\right|_{r \to 0} &=& C_3\left(3a + {e^{ - 3ar}}\right ),
\end{eqnarray}

So $\rho_0 (r)$ has maximum on the brane and falls off at the infinity as $e^{-3ar}$. Now, using \eqref{BS-OutsideMetricAnsatz} and \eqref{F}, it's easy to find the time-averaged components of the zero mode tensor
\begin{eqnarray}\label{VF-ZeroModeTensor}
\left\langle {F_t^i} \right\rangle  &=&  - {e^{ - 2ar}}{\rho _0}\left( {f_t^i + \left( {{K_1} - 1} \right){\partial ^i}{a_t}} \right)\,, \nonumber \\
\left\langle {F_i^k} \right\rangle  &=&  - {e^{ - 2ar}}{\rho _0}f_i^k\,,\,\\ \nonumber
\left\langle {F_r^\beta } \right\rangle  &=&  - {e^{ - 2ar}}\frac{{d{\rho _0}}}{{dr}}{a^\beta }\,,\\ \nonumber \left\langle {F_\theta ^A} \right\rangle  &=& 0\,,
\end{eqnarray}
where ${f_{\mu \nu }}({x^\alpha }) = {\partial _\mu }{a_\nu }({x^\alpha }) - {\partial _\nu }{a_\mu }({x^\alpha })$ and small Latin indexes $i$ and $k$ run values $x$, $y$ and $z$. Accordingly the 6D Lagrangian for zero mode is
\begin{widetext}
\begin{eqnarray}\label{VF-ZeroModeLagrangian}
\left\langle {{L_1^{(0)}}} \right\rangle  =  - \frac{1}{4}\sqrt { - g} \left\langle {F_A^B} \right\rangle \left\langle {F_B^A} \right\rangle  =  - \frac{{{R_0}}}{4}{e^{ar}}\left[ {{\rho _0}^2f_\mu ^\nu f_\nu ^\mu  + {\rho _0}^2\left( {{K_1} - 1} \right)\left( {f_t^i{\partial ^t}{a_i} + f_i^t{\partial ^i}{a_t}} \right) + } \right. \nonumber \\
 + \left. {{\rho _0}^2{{\left( {{K_1} - 1} \right)}^2}{\partial ^i}{a_t}{\partial ^t}{a_i} + {{\left( {\frac{{d{\rho _0}}}{{dr}}} \right)}^2}{a_\mu }{a^\mu }} \right]~.
\end{eqnarray}
\end{widetext}
Using \eqref{BS-K_b(r)-Asymptotics} and \eqref{VF-Solution-rho-asymptotics}, it's straightforward to show that
on the brane, $r=0$, the Lagrangian has standard 4D form

\begin{equation}\label{VF-ZeroModeLagrangian-AsymptoticsCloseToBrane}
\begin{array}{r}
{\left. {\left\langle {L_1^{(0)}} \right\rangle } \right|_{r = 0}} =  - \frac{{{R_0}}}{4}{C_3}^2{\left( {3a + 1} \right)^2}f_\mu ^\nu f_\nu ^\mu ,
\end{array}
\end{equation}
and far from the brane, $r\to \infty$, it has the following asymptotic form
\begin{equation}\label{VF-ZeroModeLagrangian-Asymptotics}
{\left. {\left\langle {L_1^{(0)}} \right\rangle } \right|_{r \to \infty }} =  - \frac{{{R_0}}}{4}{C_1}^2{e^{ - 5ar}}f_\mu ^\nu f_\nu ^\mu  ~.
\end{equation}
So, the integral over extra coordinates $r$ and $\theta$ in the action \eqref{VectorAction} is convergent, what means that 4D vector fields are localized on the brane.

\section{Localization of Spin 1/2 Fermionic Fields}
The purpose of this section is to explicitly show that there exists normalizable zero mode of the spin $1/2$ fermionic field. The starting action is the Dirac action
\begin{equation}\label{FF-DiracAction}
{S_{\frac{1}{2}}} = \int {{d^6}x\sqrt { - g} \,\bar \Psi i{\Gamma ^A}D{}_A\Psi }~ ,
\end{equation}
from which the equation of motion is
\begin{equation}\label{FF-DiracEquation}
0=\Gamma ^AD_A\Psi=\left( {{\Gamma ^\mu }{D_\mu } + {\Gamma ^r}{D_r} + {\Gamma ^\theta }{D_\theta }} \right)\Psi ~ .
\end{equation}
We introduce the vielbein through the conventional definition:
\begin{equation}\label{FF-VielbeinDefinition}
g_{AB}=\eta _{\bar A\bar B}h^{\bar A}_A h^{\bar B}_B~,
\end{equation}
where $\bar A,\bar B, ...$, refer to 6D local Lorentz (tangent) frame. Using the relation ${\Gamma ^A} = h_{\bar B}^A{\gamma ^{\bar B}}$ with $\Gamma^A$ and $\gamma^{\bar B}$ being the curved gamma matrices and the flat gamma ones, respectively, we have relations
\begin{eqnarray}\label{FF-Relation-Gamma-gamma}
  {\Gamma ^t} &=& {e^{ - ar}}{\gamma ^t},~~~{\Gamma ^i} = {e^{ - ar - u/2}}{\gamma ^i},\\ \nonumber
  {\Gamma ^r} &=& {\gamma ^r},~~~{\Gamma ^\theta } = R_0^{-1}{e^{ - ar + 3u/2}}{\gamma ^\theta },
\end{eqnarray}
 where the index $i$ runs values $x$, $y$ and $z$. The covariant derivatives in \eqref{FF-DiracAction} and \eqref{FF-DiracEquation} are
 \begin{equation}\label{FF-CovariantDerivatevs}
{D_A} = {\partial _A} + \frac{1}{4}\Omega _A^{\bar B\bar C}{\gamma _{\bar B}}{\gamma _{\bar C}}~,
 \end{equation}
 with $\Omega _A^{\bar B\bar C}$ being spin-connections.
The nonvanishing spin-connection components for our background metric \eqref{BS-OutsideMetricAnsatz} are
\begin{eqnarray}\label{FF-Spin-connection-NonvanishingComponents}
\Omega _x^{\bar t\,\bar x} &=& \Omega _y^{\bar t\,\bar y} = \Omega _z^{\bar t\,\bar z} =  - {\partial _t}\left( {{e^{u/2}}} \right), \nonumber \\
\Omega _\theta ^{\bar t\,\bar \theta } &=&  - {R_0}{\partial _t}\left( {{e^{ - 3u/2}}} \right), \nonumber \\
\Omega _t^{\bar r\,\bar t} &=& a{e^{ar}},\\\Omega _x^{\bar r\,\bar x} &=& \Omega _y^{\bar r\,\bar y} = \Omega _z^{\bar r\,\bar z} = {\partial _r}\left( {{e^{ar + u/2}}} \right), \nonumber \\ \nonumber  \Omega _\theta ^{\bar r\,\bar \theta } &=& {R_0}{\partial _r}\left( {{e^{ar - 3u/2}}} \right).
\end{eqnarray}
Taking into account these results the Dirac equation can be written as
\begin{widetext}
\begin{equation}\label{FF-DiracEquation-TimeAverage}
\left( {{\gamma ^t}{\partial _t} + {K_{\frac{1}{2}}}\left( {{\gamma ^x}{\partial _x} + {\gamma ^y}{\partial _y} + {\gamma ^z}{\partial _z}} \right) + {e^{ar}}{\gamma ^r}\left( {\frac{{5a}}{2} + {\partial _r}} \right) + \frac{1}{{{R_0}}}{K_{\frac{3}{2}}}{\gamma ^\theta }{\partial _\theta }} \right)\Psi  = 0~.
\end{equation}
\end{widetext}
We look for solutions of the form
\begin{equation}\label{FF-Spinor-Psi-decomposition}
\Psi ({x^A}) = \psi (x^\nu )\sum\limits_{l,m} {{\alpha _m}(r){e^{il\theta }}}~,
\end{equation}
 where $\psi (x^\nu )$ satisfies the massless 4D Dirac equation ${\gamma ^\mu }{\partial _\mu }\psi ({x^\nu }) = 0$. Using \eqref{FF-Spinor-Psi-decomposition}, for the $s$-wave ($l=0$) zero mode fermionic wave function Dirac equation \eqref{FF-DiracEquation-TimeAverage} reduces to
 \begin{widetext}
 \begin{equation}\label{FF-DiracEquation-ZeroMode}
 \left( {{\gamma ^r}\left( {\frac{{5a}}{2} + {\partial _r}} \right) + {e^{ - ar}}\left( {{K_{\frac{1}{2}}} - 1} \right)\left( {{\gamma ^x}{\partial _x} + {\gamma ^y}{\partial _y} + {\gamma ^z}{\partial _z}} \right)} \right)\psi ({x^\nu }){\alpha _0}(r) = 0~.
 \end{equation}
\end{widetext}
 As in previous sections, we explore (\ref{FF-DiracEquation-ZeroMode}) in the two limiting regions: far from ($r \to  + \infty $) and close to ($r\to 0$) the brane. Taking into account \eqref{BS-K_b(r)-Asymptotics}, in  both regions the equation reduces to
 \begin{equation}\label{FF-DiracEquation-ZeroMode-AtInfinity}
 \left( {\frac{{5a}}{2} + {\partial _r}} \right){\alpha _0}(r) = 0~.
 \end{equation}
So, the solution in these to regions has forms
\begin{eqnarray}
\label{FF-ZeroMode}
{\left. {{\alpha _0}(r)} \right|_{r \to 0}}{\rm{ }} = {C_1}{e^{ - 5ar/2}},\\
\nonumber {\left. {{\alpha _0}(r)} \right|_{r \to  + \infty }}{\rm{ }} = {C_2}{e^{ - 5ar/2}}~,
\end{eqnarray}
with $C_1$ and $C_2$ being some real constants. Then, using \eqref{BS-K_b(r)-Asymptotics} and \eqref{FF-ZeroMode}, in these regions the zero mode time-averaged Lagrangian $L_{1/2}^{\left( 0 \right)} = \sqrt { - g} \left\langle {\bar \Psi i{\Gamma ^A}{D_A}\Psi } \right\rangle $ has the following forms
\begin{eqnarray}
{\left. {L_{\frac{1}{2}}^{(0)}} \right|_{r = 0}} &=& {R_0}{C_1}\,\bar \psi ({x^\mu })i{\gamma ^\nu }{\partial _\nu }\psi ({x^\mu }) ~,
\\ \nonumber {\left. {L_{\frac{1}{2}}^{(0)}} \right|_{r \to  + \infty }} &=& {R_0}{C_2}{e^{ - ar}}\,\bar \psi ({x^\mu })i{\gamma ^\nu }{\partial _\nu }\psi ({x^\mu })~.
\end{eqnarray}
On the brane the 6D Lagrangian has the standard 4D form, and the integral over extra coordinates $r$ and $\theta$ in in the action \eqref{FF-DiracAction} is convergent. So the 4D fermions are localized on the brane.
\section{Localization of Gravitons}
For spin 2 gravitons we consider the following metric fluctuations
\begin{equation}
d{s^2} = {e^{2ar}}\left( {{g_{\mu \nu }} + {h_{\mu \nu }}} \right)d{x^\mu }d{x^\nu } - d{r^2} - R_0^2{e^{2ar - 3u}}d{\theta ^2}~, \nonumber
\end{equation}
where
\begin{equation}
{g_{\mu \nu }} = (1,\, - {e^u}, - {e^u},\, - {e^u})~. \nonumber
\end{equation}
It is straightforward to show that equations of motion of the fluctuations $h_{\mu \nu}$ are
\begin{equation}
\frac{1}{{\sqrt { - g} }}{\partial _M}\left( {\sqrt { - g} \,{g^{MN}}{\partial _N}{h_{\mu \nu }}} \right) = 0~. \nonumber
\end{equation}
These equations of motion for the fluctuations in the present background are equivalent to that of scalar field \eqref{SF-KGEquation1} considered in Sec. 3. Accordingly, the localization problems for spin 2 graviton and scalar field  are similar, and gravitons are also localized on the brane.
\section{Summary and Conclusions}
In the article we have introduced new nonstationary 6D standing wave braneworld model generated by gravity coupled to a phantom-like scalar field and have explicitly shown that spin $0$, $1$, $1/2$ and $2$ fields are localized on the brane by the universal and purely gravitational trapping mechanism. In our model, as opposed to earlier static approaches with decreasing warp factors \cite{Od,Oda1}, localization takes place in the case of increasing warp factor.

Finally, it should be noted that in the article we have considered the model for $\omega/a=Z_1$, where $Z_1$ is the first zero of the Bessel function $J_{\frac{5}{2}}$. But according to condition \eqref{BS-Quantization-omega-a-epsilon} there is possibility to control the number of circular islands in the 2D extra space that can provide interesting approach to the old problem - the nature of flavor.

%


\end{document}